\documentclass[article]{aa}
\usepackage{graphicx}

\begin{document}

\title{Jet rotation: launching region, angular momentum balance
and magnetic properties in the bipolar outflow from RW Aur. 
 \thanks{Based on observations made with the NASA/ESA
 {\em Hubble Space Telescope}, obtained at the Space Telescope
 Science Institute, which is operated by the Association
of Universities for Research in Astronomy,
Inc., under NASA contract NAS5-26555.}}

\author{Jens Woitas \inst{1}
        \and Francesca Bacciotti \inst{2}
        \and Thomas P. Ray \inst{3}
        \and Alessandro Marconi \inst{2}
        \and \\
        Deirdre Coffey \inst{3}
        \and Jochen Eisl\"offel \inst{1}
        }

\offprints{Jens Woitas, \email{woitas@tls-tautenburg.de}}

\institute{Th\"uringer Landessternwarte Tautenburg,
           Sternwarte 5, D-07778 Tautenburg, Germany \and
           INAF - Osservatorio Astrofisico di Arcetri, Largo E. Fermi 5,
           I-50125  Firenze, Italy \and
           School of Cosmic Physics, Dublin Institute for Advanced Studies,
           5 Merrion Square, Dublin 2, Ireland}
           
\date{Received / Accepted}

\abstract{
Using STIS on board the HST we have obtained a spectroscopic
map of the bipolar jet from \object{RW~Aur} with the slit parallel to the
jet axis and moved across the jet in steps of $0\farcs07$. After applying
a velocity correction due to uneven slit illumination
we find signatures of rotation within the first 
300 AU of the jet (1\farcs5 at the distance of RW Aur). Both lobes rotate in 
the same direction (i.e. with different helicities), with toroidal velocities
in the range 5 - 30 km s$^{-1}$ at 20 and 30 AU from the symmetry 
axis in the blueshifted  and redshifted lobes, respectively.
The sense of rotation is anti-clockwise looking from the tip of the blue lobe 
(P.A. 130$^{\circ}$  north to east) down to the star.
Rotation is more evident in the [OI] and [NII] lines and at the 
largest sampled distance from the axis. 
These results are consistent with other STIS observations carried out with
the slit perpendicular to the jet axis, and with 
theoretical simulations. 

Using current magneto-hydrodynamic models for the launch of the jets, 
we find that the mass ejected in the observed part of the outflow 
is accelerated from 
a region in the disk within about 0.5 AU
from the star for the blue lobe, and within 1.6 AU from the star 
for the red lobe.
Using also previous results we estimate upper and lower limits 
for the angular momentum transport rate of the jet. 
We find that this can  be  a large fraction
(two thirds or more) of the estimated rate transported through the relevant
portion  of the disk.
The magnetic lever arm (defined as the ratio $r_A/r_0$ between  the Alfv\`en
and footpoint radii) is in the range 3.5 - 4.6
(with an accuracy of 20 - 25 \%), 
or, alternatively,  the ejection index 
$\xi = d \ln ( \dot{M}_{\mathrm{acc}} ) / d r $ is in the range 
0.025 - 0.046 (with similar uncertainties). The derived values are in
the range predicted by the models, but they also suggest that some 
heating must be provided at the base of the flow.
 
Finally, using the general disk wind theory we derive the ratio 
$B_{\phi} / B_p$ of the toroidal and poloidal components of 
the magnetic field at the observed location (i.e. about 80 - 100 AU 
above the disk).
We find this quantity to be  3.8 $\pm$ 1.1  at 30 AU from the axis in the
red lobe and -8.9 $\pm$ 2.7  at 20 AU from the axis 
in the blue lobe (assuming cylindrical coordinates centred on the star and 
with positive $z$ along the blue lobe). 
The toroidal component appears to be dominant, which would be consistent 
with  magnetic collimation of the jet.  The field appears 
to be more tightly wrapped on the blue side.

}

\titlerunning{Jet rotation and derived physical parameters in the RW Aur jet}
\authorrunning{Woitas et al.}
\maketitle

\section{Introduction} \label{intro}

The collimated Herbig-Haro (HH) jets observed 
on parsec-scale lengths in  star formation regions 
are always associated with young stellar objects (YSOs) 
that are still in their 
accretion phase. Accretion and ejection of matter are  
believed to be intimately  related phenomena, 
through the  presence of a magnetized accretion disk.
The complex interplay between accretion and ejection is modelled using
several theoretical approaches (cf. Camenzind et al.\,\cite{Cam90}, Ferreira
\cite{Fer97}, K\"onigl \& Pudritz \cite{Koe00}, Shu et al.\,\cite{Shu00},
and references therein). These models have in common the idea that
the jet is generated through the interaction of plasma with rotating
magnetic field lines that are anchored to the star/disk system.
The fluid particles lifted from the disk are forced to slide
along the rotating magnetic field lines, and are 
accelerated and collimated into bipolar jets.
The most attractive aspect  of this approach is the fact that the 
magneto-centrifugal scenario at the same time justifies 
the acceleration of the jets and the extraction of the excess angular 
momentum from the disk. This mechanism,
in combination with disk viscosity (the nature of which, however,
is still uncertain), should  contribute to slow the disk material
down to sub-Keplerian rotation. In this way matter 
is allowed to move radially toward the central star and finally
accrete onto it. 

This compelling theoretical picture, however, has received little
in the way of observational confirmation, as pointed out by
Eisl\"offel et al. (\cite{Eis00}).
The main reason is that the process occurs at small distances, corresponding,
for the nearest star formation regions (130 - 150 pc), to subarcsecond
scales.
 
The ``core'' of the central engine lies within a few AU from the source,
which is not spatially resolvable with current observational techniques. 
Important constraints on the launching region, however, can be deduced 
from the quantities observed in  the first 100-200 AU of the flow,
in which the jet achieves its collimation 
and  its final poloidal velocity. This region 
has been accessed recently with high angular resolution observations 
of jets from evolved T Tauri stars.
Such objects have been observed from space  with the 
Hubble Space Telescope Imaging Spectrograph 
(HST/STIS, 0\farcs1 resolution, 
e.g. Bacciotti et al. \cite{Bac00},  Woitas et al. \cite{Woi02}), and from 
the ground using large telescopes with Adaptive Optics 
(e.g. Dougados et al.\, \cite{Dou00}).  

Among the properties investigated in the above studies, a special place 
is taken by the detection and analysis of the {\em rotation} of the jets 
around their symmetry axis (Bacciotti et al. \cite{Bac02}, 
Testi et al. \cite{Tes02}, 
Anderson et al. \cite{And03}, Coffey et al. \cite{Cof04}, 
Pesenti et al. \cite{Pes04}).
These studies have shown that the measured toroidal 
velocities are similar to the values  predicted by the
disk-wind magneto-centrifugal models. 
The obtained measures of rotation, however, also help to constrain
the properties of the accretion/ejection ``machine''. 
For example, in combination with mass density estimates from line ratios, 
rotation profiles can yield the amount of angular momentum carried away 
from the  accreting system by the jet. This can then 
be compared with the disk accretion properties. 
As another example, the application of special conservation 
laws can in principle give information on the 
properties of the magnetic field in the jet,
a quantity that is fundamental to all jet acceleration models, but  
that is notoriously difficult to directly examine by observation.

The aim of this paper is to illustrate the potential offered 
by such a combined observational/theoretical investigation to
constrain the accretion/ejection structure at the origin of the jet. 
First, we present results from a new observational study 
of the RW Aur jet, conducted with multiple slits
oriented along the flow axis (Sections 2 and 3). 
The adopted technique is similar to the one used in our first rotation
study, that concerned the jet from \object{DG Tauri} (Bacciotti et al. 
\cite{Bac02}).
Here, however, the technique is applied for the 
first time to a {\em bipolar} jet, on a more extended and more finely 
sampled region. In addition, more details are given about the 
correction routines for  uneven slit illumination.
Then, in the second part of the paper (Section 4), 
we combine the  obtained results with morphological 
and excitation properties of 
this jet derived on subarcsecond scales from previous studies and
partly based on the same dataset
(Woitas et al. \cite{Woi02} , hereafter Paper I,  
Dougados et al. \cite{Dou00}).
In this way we show how, using general theoretical principles,  
one can derive physical quantities that are crucial
for a description of the accretion/ejection region close to the origin
of the jet. Finally, in Section 5 we summarize our conclusions.

\begin{figure}
\resizebox{\hsize}{!}{\includegraphics[width=12cm]{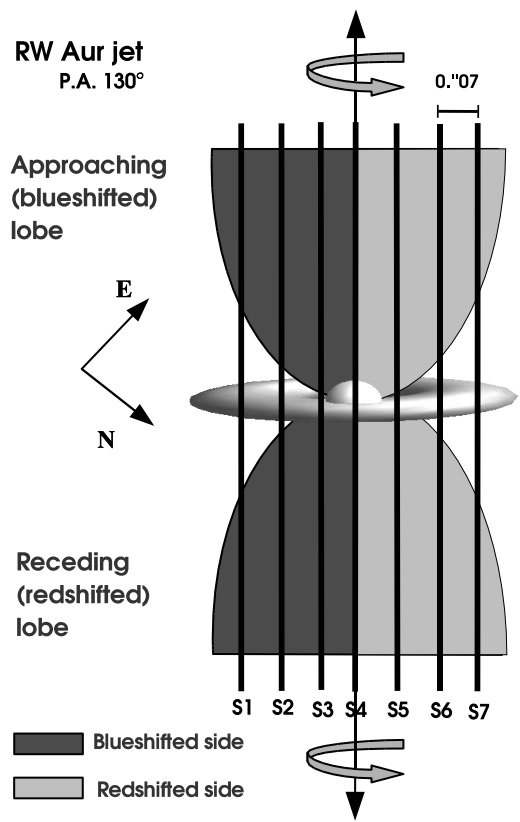}}
\caption{\label{slitconfig}
Schematic drawing of the observing mode for the bi-polar jet from RW Aur.
The STIS slit is kept parallel to the flow axis, 
and stepped sequentially in seven different positions 
(labelled $S1$, $S2$....$S7$) each time by 0\farcs07.
The slit width is 0\farcs1, so the apertures are slightly overlapping.
The sense of rotation derived in the present paper is illustrated
by the arrows around the symmetry axis.}
\end{figure}

\section{Observations and data analysis} \label{obs}
The HST/STIS observations of RW Aur were carried out on 10 December
2000. Details of the data acquisition and reduction were described
in Paper~I. Briefly, the dataset consists of seven spectra (hereafter
$S1$,...,$S7$) taken at slit positions parallel to the jet axis
(P.A.~=~$130^{\circ}$, Dougados et al.\,\cite{Dou00}).
This observing procedure is shown in Fig.\,\ref{slitconfig}.
The pixel scale is $0\farcs05/\mathrm{pixel}$ for the
spatial axis and $0.554\,\mathrm{{\AA}}/\mathrm{pixel}$ for the
dispersion axis. The latter value corresponds to
$\approx 25~\mathrm{km}\,\mathrm{s}^{-1}$/pixel.
The slit width is $0\farcs1$.
The covered spectral range contains H$\alpha$ and the prominent forbidden
emission lines [OI]$\lambda\lambda$6300,6363, [NII]$\lambda\lambda$6548,6583
and [SII]$\lambda\lambda$6716,6731. 
However, blueshifted [OI]$\lambda$6300 emission
cannot be used for the present analysis because the line profile
is cut by the detector edge.\\
In the course of the observations the slit is moved transversely
across the jet in steps of $0\farcs07$ from southwest ($S1$) to northeast
($S7$). The position of $S4$ corresponds to the jet axis. If the jet
rotates, one expects non-zero peak velocity differences $S7 - S1$,
$S6 - S2$ and $S5 - S3$ that have the same sign at all separations
from the star. However, one has to
take into account that uneven slit illumination may introduce spurious
velocity shifts causing peak velocity differences similar to those
described above, which may then be present also in the case of a
non-rotating jet.
This effect has been eliminated applying correction 
routines developed by one of us (A.M.),
which also have been used to study the rotational motions in the DG Tau 
jet. The calculations performed in the correction routine are illustrated
in Appendix A of Bacciotti et al.\,(\cite{Bac02}) and in Marconi
et al.\,(\cite{Mar03}), and the procedure can be briefly summarized in
three steps as follows. 

(i)
In the first step a bidimensional surface $I_{\mathrm{\lambda}}(x,y)$ 
is calculated that simulates the 
{\em observed} brightness distribution in each emission line at the positions
(x,y) on the sky intercepted by the seven STIS slits.
     This is done by fitting separately for each line a series 
of 2-D surfaces with an elliptical base 
simulating the knots along the flow to the 
line fluxes  measured along the seven slits at once. The best 
fit is found through an iterative algorithm.
     It is then necessary to assume a model radial velocity field of the nebula
$u(x,y)$. In our case we wish to determine the spurious velocity 
caused by STIS, so we assume an arbitrary constant value $\bar{u}$, equal 
for all the slits and  positions on the sky, 
i.e. a non-rotating uniformly moving jet is assumed. 

(ii)
In the second step the routines calculate the average velocity measured 
at the detector $v_{\lambda}(x,y)$ for each slit, line and position
considered.
This is done by convolving the model velocity field as it would be seen 
at the detector without optical distortions with the  model of the 
brightness distribution obtained at step (i), 
with the Point Spread Function of HST at the 
wavelength of interest and with the slit aperture offset with respect to the
axis of the jet  as in the observations. Details of this 
calculation can be found in the Appendix of Bacciotti et al. (\cite{Bac02}). 
Since the model velocity of the gas is set to be the {\em same} value for 
all the slits, the searched instrumental spurious velocity is 
$V_{\lambda}(x,y) = v_{\lambda}(x,y) - \bar{u}$, the result being 
{\em independent} of the chosen value of the model radial velocity. 
The absolute values of $V_{\lambda}(x,y)$ 
range between 2 and 8 km s$^{-1}$ and 
have opposite sign depending on the relative position of the slit and 
the jet axis, since slits to the left or to the right of the axis have 
opposite illumination gradients. 
The values for $V_{\lambda}(x,y)$ are higher (in absolute value) where 
the illumination gradient is steeper across the jet, i.e. typically 
at the position of the knot peaks.

(iii)  Finally, the spurious ``shift'' is 
calculated between the considered opposed slits as,
e.g. $V_{\lambda}(S7,y) - V_{\lambda} (S1,y)$.
The  calculated shifts are instrumental offsets, 
and therefore they are {\em subtracted} from the raw 
shift measurements, to cancel the effect 
of the uneven slit illumination.

The  spurious velocity depends only on the gradient of the slit illumination,
and not on the integrated flux across the slit,  
so any systematic error in the photometry would not affect our determination. 
Therefore, the uncertainty associated with the correction 
depends only on the accuracy with which the {\em  
shape}  of the  observed emission
distribution is fitted by the assumed 2-D surfaces convolved with 
the instrumental response. 
The other parameters entering
the calculation (see Bacciotti et al. \cite{Bac02}), i.e. 
the instrument PSF, the 
detector scale factor and  the line flux measured by 
STIS are all known with a much higher accuracy than that associated 
with the fitting procedure. 

The accuracy of the correction has been estimated 
{\em a posteriori} by investigating the variations in the determined 
spurious shifts  due to an imposed change in the scale factor of the 
brightness distribution of the surface fitting each knot. 
Variations of the adopted illumination gradient producing a 
change by 10 - 15 \% in the value of the fitted flux, 
which is about the accuracy of our iterative fitting algorithm, lead to  
a change of 5 - 8\% in the determination of the spurious velocity shift.
Thus we take this factor as the uncertainty of the 
correction, which, in the case of the largest spurious shifts 
($\sim 16 $ km s$^{-1}$), corresponds to less 
than $\pm$ 1.5 km  s$^{-1}$. 
Also, we checked that imposed  small misplacements of the slit 
positions or inclination (up to 10\%) do not lead to any appreciable 
change in the determination of the spurious velocity. 
 
It should be noted that in the real case in which the velocity 
field is not constant, an additional deformation is introduced by the passage 
of the light through the telescope and spectrograph. In our case, however, 
this contribution can be neglected because we are actually interested in 
spurious velocity {\em differences} between symmetrically opposed slit pairs, 
and the additional contribution would cancel in the difference 
because the absolute value of the real radial velocity, 
as well as the intensity, have a symmetrical variation with respect to the 
jet axis or central slit.

In practice, the spurious velocity is positive for the slits 
$S1$, $S2$ and $S3$, located south-west of the axis, and negative 
for $S5$, $S6$, $S7$. So for the adopted configuration
($S7 - S1$, $S6 - S2$, $S5 - S3$)
the instrument produces  a spurious  negative {\em shift} in both jet lobes. 
The effect is more pronounced for the forbidden lines (FELs) than it is for
H$\alpha$. This is due to the fact that the bright 
H$\alpha$ emission is more evenly distributed across the transverse 
direction in the area covered by the seven slits (see Paper~I).
Similarly, other detected differences in the spurious shifts for the different 
lines can be attributed to the different 
transverse spatial distribution of the emission (see Table 2 and
its description in Section 3).  

After subtracting the spurious negative shifts the rotation signature
becomes more evident  for the outermost slit pairs (larger positive shift),
while in the innermost pair the measured raw negative shifts are turned back
to weakly positive or null (i.e. consistent with zero inside the measurement 
error of $\pm$ 7 km s$^{-1}$). In some positions of the innermost pairs, 
however, the net shift in [SII] and H$\alpha$ lines remains significantly 
negative. For an explanation see Section 3.

\begin{figure}
\resizebox{\hsize}{!}{\includegraphics{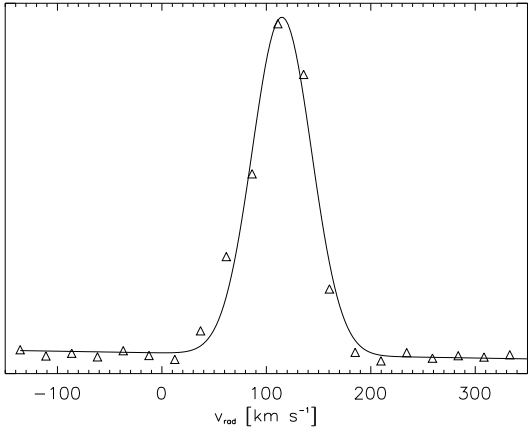}}
\caption{\label{gaussfit} Example of a Gaussian fit to the line
 profile that yields the peak velocity. This plot shows
 [SII]$\lambda$6731 at slit position S3 and at a separation of
 $0\farcs9$ from the star.}
\end{figure}

The validity of the
illumination correction can be tested using the H$\alpha$ emission
at the stellar position. As this emission is close to saturation, the
line profiles will be dominated by the HST PSF and are not supposed
to show any rotation signatures. Any velocity shifts between lateral
slits will be due to instrumental effects. The finding that there
are no significant velocity shifts in H$\alpha$ close to the star
{\it after} applying the illumination correction (see Fig.\,\ref{rw-red})
thus strongly indicates that this correction routine really removes velocity
shifts caused by uneven slit illumination. \\
The peak velocities of individual lines in all spectra and in both
outflow lobes were determined from Gaussian fits to the line profiles.
Many T~Tauri jets show two velocity components of forbidden line
emission, a high velocity component (HVC) with
$v_r\approx 100 - 200\,\mathrm{km}\,\mathrm{s}^{-1}$ and an additional
low velocity component (LVC) with a typical radial velocity of
$5$ to $20\,\mathrm{km}\,\mathrm{s}^{-1}$ (Hirth et al. \cite{Hir97}).
The HVC is thought to come from a collimated jet close to the star,
while the LVC might be the signature of a poorly collimated 
disk wind (Kwan \& Tademaru \cite{Kwan88}, \cite{Kwan95}). In Paper I
we have, however, demonstrated that there is no significant
separate LVC in our data. Therefore fitting single Gaussians is sufficient.
As an example we show in Fig.\,\ref{gaussfit} the fit to the
[SII]$\lambda$6731 line at slit position $S3$ and separation $0\farcs9$
in the redshifted lobe. The uncertainty of the peak velocities
is about $\pm 5\,\mathrm{km}\,\mathrm{s}^{-1}$, and the uncertainties
of velocity {\it differences} are thus $\pm 7\,\mathrm{km}\,\mathrm{s}^{-1}$.

For some selected positions and emission lines we have derived
velocity differences also by means of a cross-correlation technique
that directly measures the displacement of two line profiles from
different slits, with a typical accuracy of $5\,\mathrm{km}\,\mathrm{s}^{-1}$
on the velocity {\em difference}. This method is in principle more
robust than Gaussian fitting as no special shape of the line
profile has to be assumed. Its usefulness is limited, however,
in situations where the velocity range is only marginally sampled,
and this is the case here with the narrow emission lines of the
RW~Aur jet. Thus cross-correlation does not lead to a distinct
improvement, but the results are consistent with those obtained from
the Gaussian fits.
Reliable measurements are not possible for all slit positions and
separations from the source. The RW Aur jet has a FWHM of about
20~AU within its first arcsecond and is thus more strongly collimated
than the jet of DG~Tau (FWHM~$\approx$ 50~AU at $d = 0\farcs5$, see Paper~I,
Fig. 4 therein). For this reason, there are cases where there is no
significant signal at distinct separations in $S1$ and $S7$. In [NII],
which traces emission very close to the jet axis, this will even be the case
for more central slit positions. Furthermore, quenching effects make the FELs
extremely weak within 0\farcs1 - 0\farcs2 from the origin, thus
we exclude this region from the analysis. Although the jet is clearly visible
up to $\approx 4''$ ($\approx~800$~AU) from the star in the STIS data
(Paper~I) we restrict the rotation analysis to the first
$\approx 300\,\mathrm{AU}$ approximately. Over this separation range
our data will reflect signatures from the initial jet channel,
whereas further away interaction between the jet and the circumstellar
environment might become important. To obtain de-projected separations
we assume the inclination angle of the RW~Aur jet to be $i = 46\pm 3^{\circ}$
with respect to the line of sight (L\'opez-Mart\'{\i}n et al.
\cite{Lop03}). In Paper~I we adopted $i = 53^{\circ}$, but this
estimate was based on the proper motion of only one jet knot
and furthermore affected by an erroneous pixel scale given by
Dougados et al.\,(\cite{Dou00}) for their 1997 observations (the corrected
value is given by L\'opez-Mart\'{\i}n et al. \cite{Lop03}).
With the new value of the inclination angle, 1\farcs5 on the sky corresponds
to 300~AU along the outflow direction.

\begin{figure*}
\centering
\includegraphics[width=16cm]{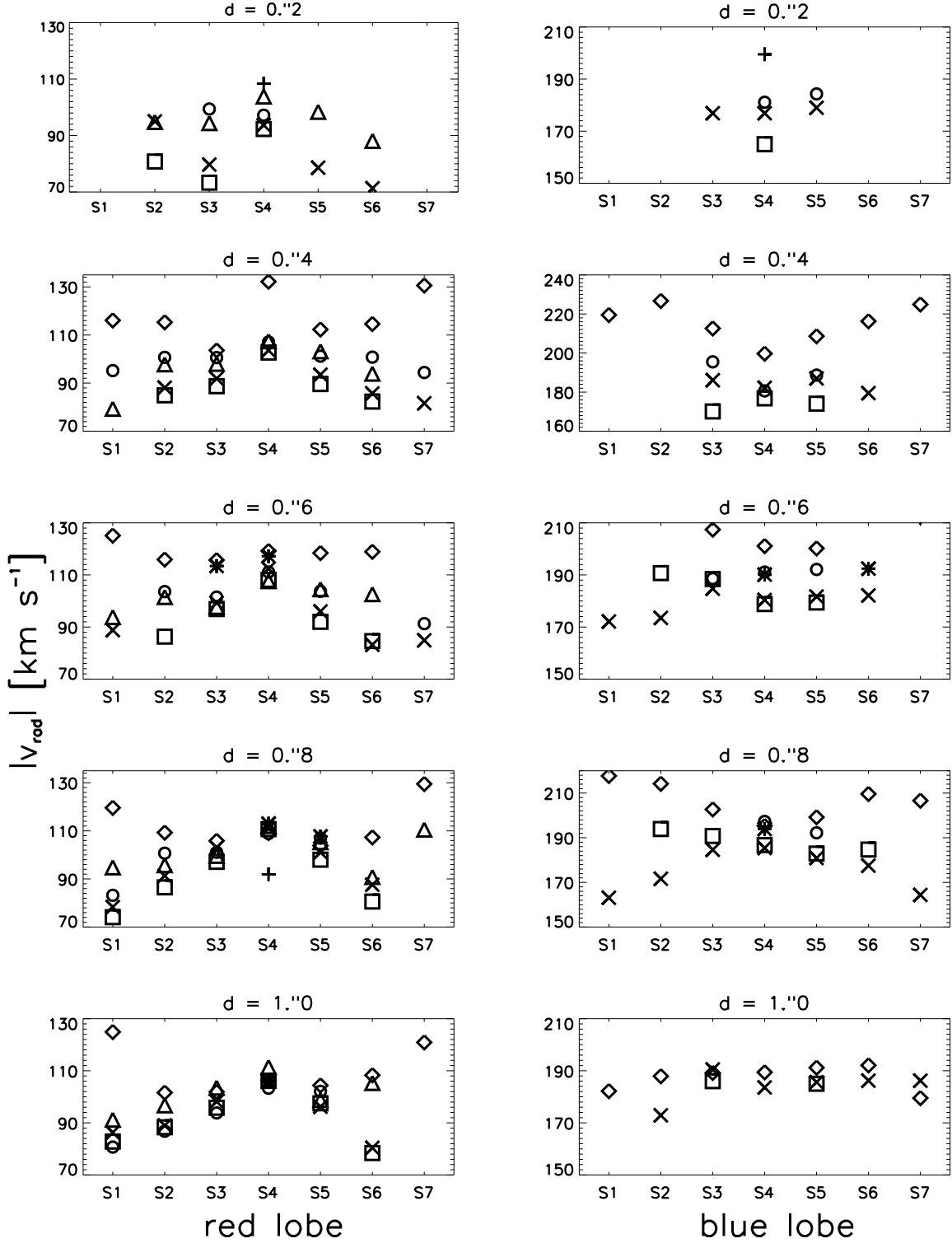}
\caption{\label{velplot} Peak velocities for different slit positions
 and separations from the star. Crosses denote [SII]$\lambda$6731,
 squares [SII]$\lambda$6716, circles [OI]$\lambda$6363, triangles
 [OI]$\lambda$6300, plus signs [NII]$\lambda$6548, asterisks
 [NII]$\lambda$6583, and diamonds H$\alpha$.}
\end{figure*}

\section{Indications of rotation} \label{res}
Fig.\,\ref{velplot} shows the peak velocities for all slit positions
and at different separations from the star. These velocities are
corrected for uneven slit illumination (see Sect.\,\ref{obs})
and for the heliocentric radial velocity of RW~Aur~A, i.\,e.
$+23.5\,\mathrm{km}\,\mathrm{s}^{-1}$ as derived from the Li\,I line at 
$\lambda\approx 6707 \mathrm{\AA}$ from our data.
We note that Petrov et al. (\cite{Pet01}) repeatedly observed
RW~Aur using high resolution echelle spectroscopy and found a
mean heliocentric $v_r$ of $+16\,\mathrm{km}\,\mathrm{s}^{-1}$
with periodic variations of $\pm~6~\mathrm{km}\,\mathrm{s}^{-1}$
($P\approx 2.8\,\mathrm{d}$). Such variation in the adopted systemic
velocity does not affect our determinations of velocity differences
between the two sides of the flow. For the present analysis, however, in which
we also give transverse profiles of the absolute radial velocity,
we decided to adopt the former value for self-consistency.\\
\begin{figure}
\resizebox{\hsize}{!}{\includegraphics{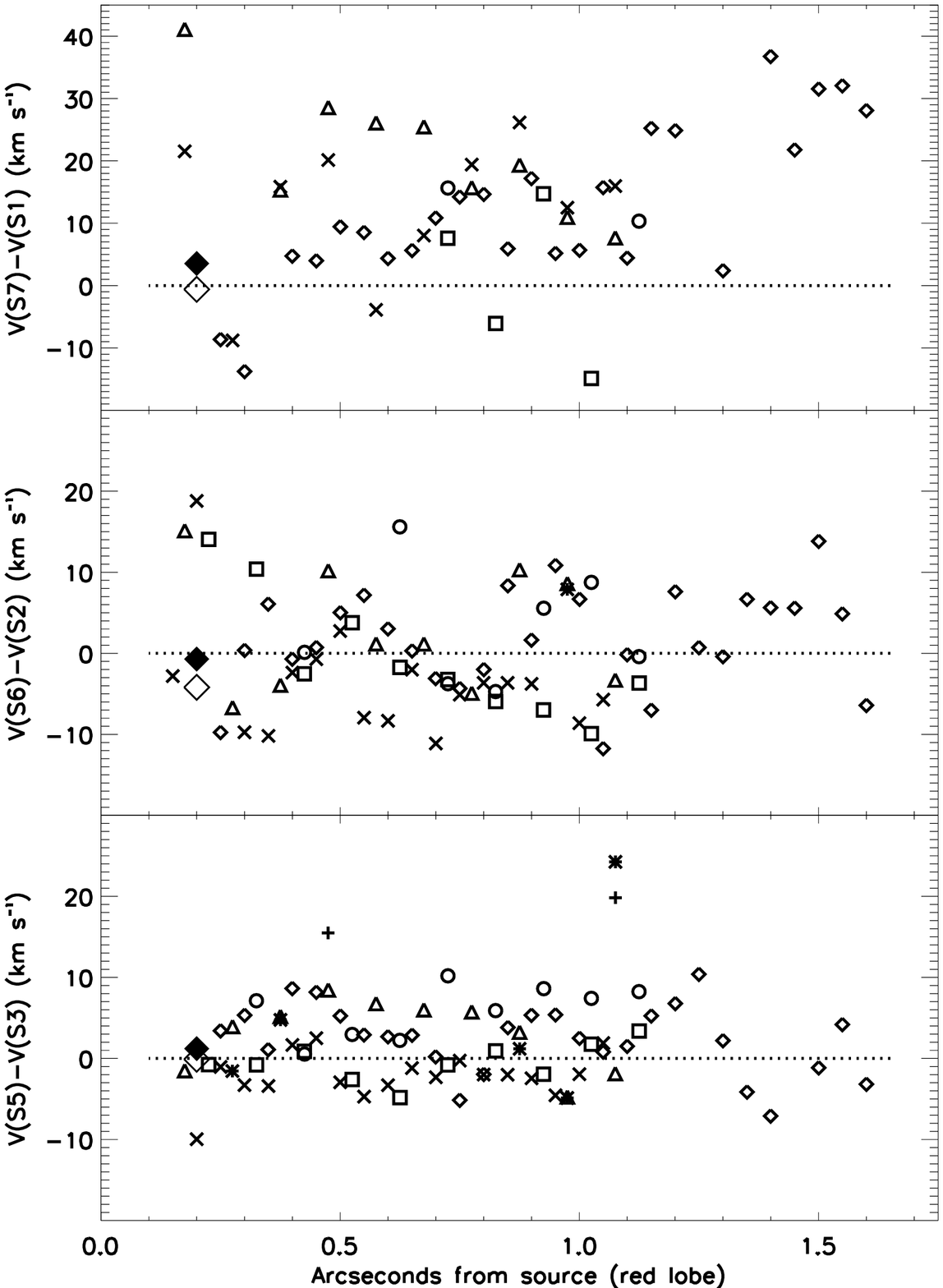}}
\caption{\label{rw-red}
Peak radial velocity differences in the redshifted outflow lobe as a 
function of separation from the source.
Upper panel: difference between the values measured at slits $S7$ and $S1$, 
or at 30~AU from the axis. Middle panel: $S6$ - $S2$, corresponding to 
20~AU  from the axis. Lower panel: $S5$ - $S3$, or 10~AU from the axis. 
 The plot symbols have the same meaning as in Fig.\,\ref{velplot}.
 The large diamonds represent H$\alpha$ at the stellar position, the open
 one before the correction for uneven slit illumination and the filled one
 afterwards.}
\end{figure}
\begin{figure}
\resizebox{\hsize}{!}{\includegraphics{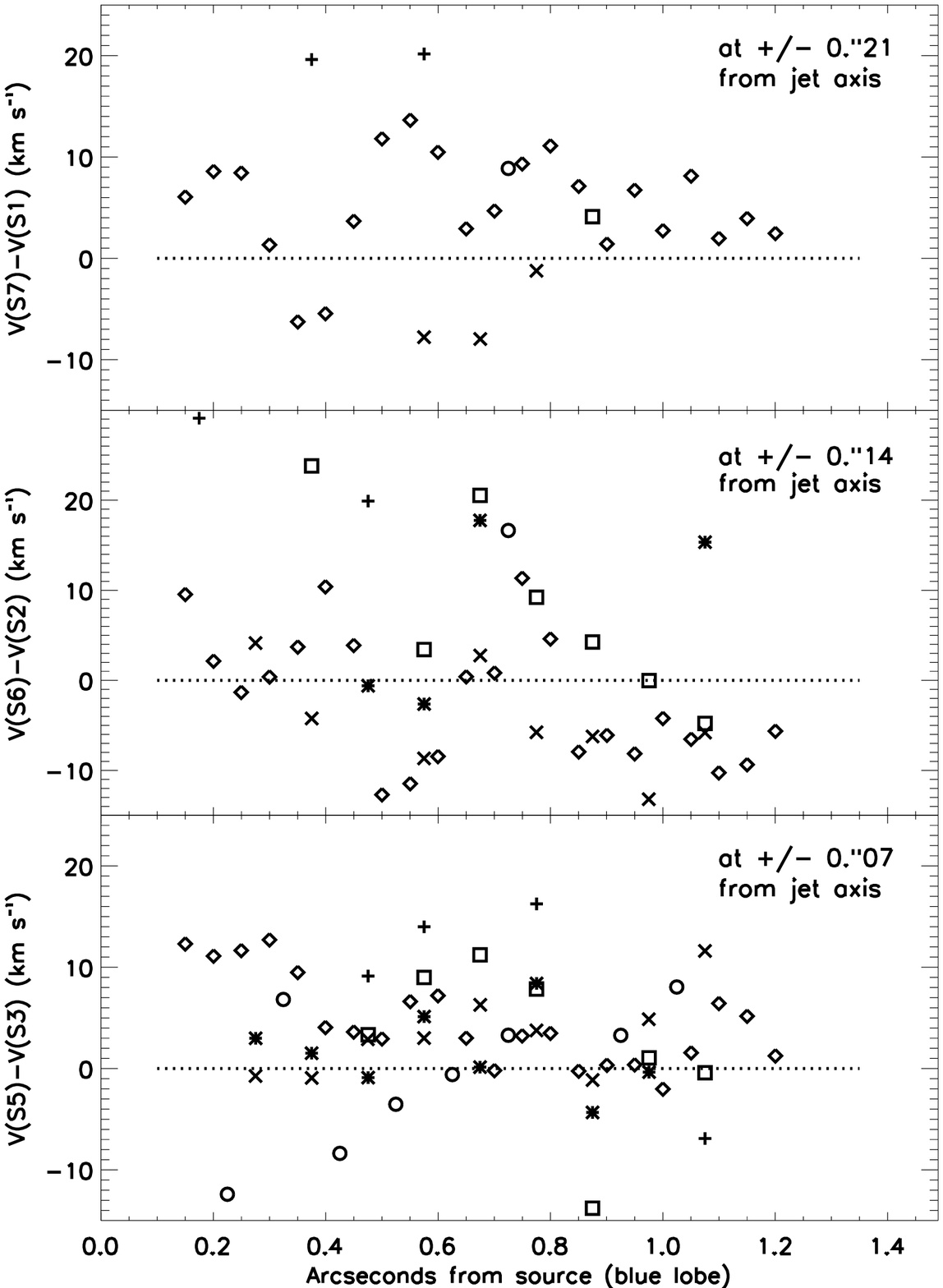}}
\caption{\label{rw-blue} Same as Fig.\,\ref{rw-red}, but for the
 blueshifted outflow lobe.}
\end{figure}
The resulting velocity differences $S7 - S1$, $S6 - S2$ and
$S5 - S3$ are plotted in Figs.\,\ref{rw-red} and \ref{rw-blue}
as a function of the separation from the star.
The result is not as impressive as for 
DG Tau, due  to the fact that the RW Aur jet is more 
collimated (see below). 
One can see, however, that in all slit pairs there
is a definite tendency for the velocity shifts to be located on the 
same side of the zero line in both lobes. The  scatter
of the data points in the different lines is sometimes large, and  
it is probably  due to different excitation conditions, as  
discussed later. Nevertheless, the averages of all velocity shifts 
in all lines and separations from the star
show  the same sign in all slit pairs and in both lobes (Table 1).
Thus, the emission from the north-eastern parts of
both lobes is more redshifted than the emission from 
the south-western parts. This leads us to 
suggest that both outflow lobes rotate clockwise 
looking from the tip of the redshifted lobe 
down to the star. The matter thus flows in the jet with 
opposite helicities. These results are fully 
consistent with the study in Coffey et al. (2004).

\begin{table}
\caption{\label{mean_veldiff} Mean velocity differences (in km\,s$^{-1}$),
 averaged over all lines and all separations from the star}
\begin{tabular}{llll}
\hline\hline
 & $S7-S1$ & $S6-S2$ & $S5-S3$ \\
\hline
Redshifted lobe   & $12.9\pm 1.7$ & $0.2 \pm 0.8$ & $2.0\pm 0.6$ \\
Blueshifted lobe  & $5.2\pm 1.3$  & $1.1\pm 1.6$ & $3.8\pm 0.9$ \\
\hline
\end{tabular}
\end{table} 

Given the fact that the measured rotation velocities are of
the order of only $\approx 10\,\mathrm{km}\,\mathrm{s}^{-1}$
one has to look carefully at artifacts that might mimic rotation.
The problem of uneven slit illumination has already been
discussed in Sect.\,\ref{obs}. Another issue that has to be
taken into account here is that the slit orientation might
not be exactly parallel to the jet axis.
Gaussians fits to the emission profiles transverse to the jet, 
as well as the output of the illumination correction routines,
do indeed show that the slit set is well-centred on the star, but the chosen
slit P.A. (130 deg) turns out to be larger by 1-2 degrees than the 
real P.A. of the jet direction (blue lobe - the same misalignment is also 
noted in Coffey et al. \cite{Cof04} ). 
This causes the slits of one given ``pair''
to trace gas with different poloidal velocity, which will lead to apparent
velocity shifts even if the jet is non-rotating. 
From Fig. 3 one can estimate that the corresponding ``false'' 
shifts would be 1 - 5 km\,s$^{-1}$, which is not negligible.
For the given position angle the mimicked
rotation would, however, have the {\em opposite} direction in comparison
to the observed motion. This means that the true rotation
velocities are in fact somewhat higher than those suggested by Figs.
\ref{rw-red} and \ref{rw-blue}, and given in Table\,\ref{mean_veldiff},
strengthening our result even further.\\
Finally, the observed velocity shifts could in principle reflect a situation
where the poloidal velocity field is not symmetric to the jet axis,
but without any rotation. This can  happen in asymmetric bow shock
wings, but is very unlikely to occur along a distance of 300~AU,
which is much larger than the knots of the RW~Aur small-scale jet
(Paper~I).\\ 
\begin{table}
\caption{\label{velshift_lines} Velocity shifts (in km\,s$^{-1}$) in
 different emission lines, averaged over all separations from the star.}
\begin{tabular}{llll}
\hline\hline
Line & $S7 - S1$ & $S6 - S2$ & $S5 - S3$ \\
\hline
{\bf Redshifted lobe} & & & \\
$\mathrm{[OI]}\lambda$6300  & 21.1 & 2.8 & 3.1 \\
$\mathrm{[OI]}\lambda$6363  & 13.0 & 3.0 & 5.9 \\
$\mathrm{[NII]}\lambda$6548 &      &     & 17.7 \\
H$\alpha$                   & 12.1 & 1.8 & 2.4 \\
$\mathrm{[NII]}\lambda$6583 &      & 7.9 & 4.8 \\
$\mathrm{[SII]}\lambda$6716 & 0.3  & -0.6 & -0.5 \\
$\mathrm{[SII]}\lambda$6731 & 12.7 & -3.8 & -2.2 \\
{\bf Blueshifted lobe} & & & \\
$\mathrm{[OI]}\lambda$6363  &  8.8   & 16.7 & 3.2 \\
$\mathrm{[NII]}\lambda$6548 & 19.9   & 24.5 & 8.1 \\
H$\alpha$                   &  5.2   & -2.0 & 4.7 \\
$\mathrm{[NII]}\lambda$6583 &        &  7.5 & 1.6 \\
$\mathrm{[SII]}\lambda$6716 &  4.1   &  8.1 & 2.6 \\
$\mathrm{[SII]}\lambda$6731 &  -5.6  & -6.4 & 3.3 \\ 
\hline
\end{tabular}
\end{table}
As mentioned above, the data show a notable scatter in Figs. 4 and 5,
and there are several possible explanations for this.
Clumpiness, spatially unresolved shocks and variations of the 
poloidal velocity may play a role in producing the observed scatter.
Some scatter would be present, however, even if the flow was laminar with 
the poloidal velocity smoothly decreasing
from the axis to the jet borders (``onion-like'' kinematic
structure), as predicted by the MHD stationary models. 
According to these models
the layers located closer to the axis rotate faster, so different
emission lines that trace different regions
along the line of sight will also show different radial velocity shifts if
measured at the same position on the sky. From Table \ref{velshift_lines}
one can see that the rotation signatures are more pronounced
in the [OI] and [NII] lines than in [SII] and H$\alpha$. We 
have indeed found that the transverse FWHM of the emission in 
[OI] and [NII] is smaller than
the FWHM in [SII] (Fig.\,4 of Paper~I). The emission in H$\alpha$ also 
presents a larger transverse extension.  
Thus at the same position across the jet, corresponding to two coupled slits, 
the ``[OI] jet''and the ``[NII] jet'' are 
observed closer to their borders than the [SII] (and H$\alpha$)
jet. Pesenti et al. (\cite{Pes04})  note that projection of the emission 
along a line of sight that crosses the beam of the jet 
tends to cancel out the radial velocity differences produced by rotation,
this effect being more severe for distances closer to the axis and
lower spatial resolution of the instrument.
Therefore, at any point of observation,
the rotation signatures will be more evident in [OI] and [NII], 
as the emission is less averaged across the line of sight 
in this case. The same projection effects 
can explain the fact that the evidence for rotation is weaker
in the innermost slit pairs than it is in $S7-S1$.
Note also that some values are negative in Table 2 for [SII] and H$\alpha$.
We interpret this as follows: We have seen above that the set of slits
is subject to an unwanted inclination of about 2 degrees 
with respect to the jet axis, in such a way that the observation
of a non-rotating jet would produce a 'false' negative shift.
On the other hand, we have pointed out above that 
for any  given slit pair the ``[OI] jet'' will show larger 
rotation shifts than the ``[SII] jet'', because of projection effects 
combined with a different spatial distribution of the emission. 
As explained above, in [OI] the rotation signature
dominates over this inclination effect.
In [SII] lines, instead, the rotation signature is very weak at the
same position. This is because the aforementioned enhanced projection
effects for the [SII] lines overwhelm the rotation here, and
the net effect is the observation of  a negative shift in [SII].
H$\alpha$ may also be affected in the innermost slits.
We have not attempted to correct for this effect, however, since the
value of the ``false'' rotation depends on the kinematic model 
assumed for the jet.
Pesenti et al. (\cite{Pes04}) also predict rotational velocity shifts 
for ``cold'' and ``warm'' 
disk wind solutions. Our observed rotation velocities are intermediate
between these two predictions, but closer to the 
warm disk wind solution. This point will be discussed further in Section~4. 

Finally, we note that a decrease of the velocity shift with distance
from the star is marginally seen in some of the lines and slit separations, 
as, for example, in the redshifted lobe for [OI]$\lambda$6300 
in the pair $S7$ -- $S1$ and $S6$ -- $S2$, and for [SII]$\lambda$6716   
in the $S6$ -- $S2$ pair, while for the blueshifted lobe 
a decrease is apparent in the H$\alpha$ points for the $S7$ -- $S1$ pair and
for [SII]$\lambda$6716 in $S6$ -- $S2$. We note that
such a behaviour is predicted by all the models of magneto-centrifugal
acceleration as a consequence of the conservation of kinetic  
angular momentum at large distance
from the Alfv\`en surface (see below) and the widening of the flow. 
Given the scatter of our datapoints and their expected accuracy, 
however, we refrain from presenting a quantitative analysis of this aspect.
Instead, in the rest of the paper we will use the  
radial velocity shifts measured in the first 0\farcs6 - 0\farcs8 
from the star to derive as much information as possible
about the physical properties of the launching region of the jet.

\section{Properties of the  accretion/ejection structure} 

From our results one can derive quantities useful to  
constrain the properties of the jet acceleration  region.
In particular it can be checked if the values obtained are consistent with 
the self-similar disk-wind models  described, e.\,g. in
K\"onigl \& Pudritz (\cite{Koe00}) and  Ferreira (\cite{Fer02}). 

\begin{figure}
\begin{center}
\includegraphics[scale=0.3]{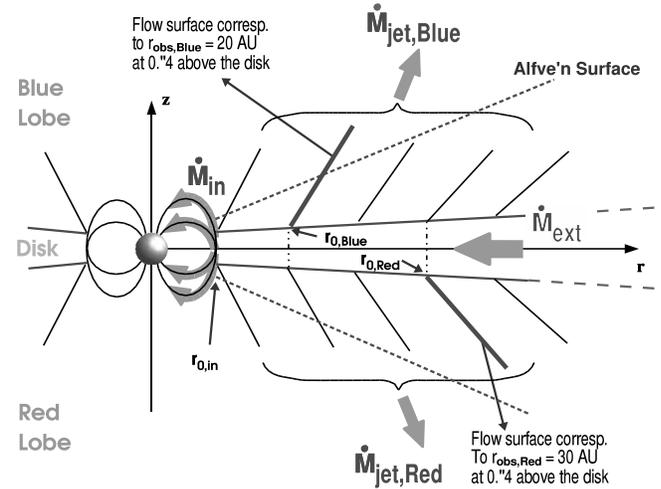}
\caption{Sketch (not to scale) of the 
configuration of the flow/magnetic  surfaces (solid thin lines)  
attached to the inner section of the accretion disk around RW Aur A, 
according to the disk-wind scenario. Thick grey arrows indicate 
the mass flows. Note the different distances of the footpoints for the
blue and red flows seen at 0\farcs4 above the disk. This is due to the
asymmetry in the bipolar jet.}
\label{diskjet}
\end{center}
\end{figure}

The launching region of the jet
is sketched  in  Fig.~\ref{diskjet}. This figure 
illustrates the configuration of the nested 
magnetic surfaces attached to the 
inner portion of the accretion disk.    
The star is at the origin of a cylindrical 
coordinate system ($r, z, \phi$), 
and the flow is assumed to be steady-state, axisymmetric and to
satisfy the ideal MHD equations. 
The poloidal component 
of the magnetic field  ${\bf B_p } = B_r {\bf e_r} + 
B_z {\bf e_z} $ is described by 
${\bf B_p } = {\bf \nabla } a \times {\bf e_\phi}/r$, where   $a(r,z) = Const$
labels the magnetic surfaces, i.e. the surfaces enclosing a constant
magnetic flux. The poloidal velocity and magnetic field are easily
shown to be parallel, and they are related by 
the expression:
$\rho {\bf v_p } = k(a) {\bf B_p }$, with
$k$ constant along each magnetic surface.  
So the magnetic and flow surfaces are coincident and can 
also be labelled with their ``footpoint'' radius $r = r_0$, or the distance 
from the star on  the  disk where the surface 
is anchored. The mass in the visible jet  is ejected from a region of 
the disk ranging from an inner radius $r_{0,\mathrm{in}}$ to
some outer radius $r_{0,\mathrm{ext}}$ to be determined.
The relevant mass fluxes are denoted by large grey arrows in the figure.

The disk and the rigidly anchored field lines rotate 
rapidly, and the centrifugal force  makes the fluid 
parcels lifted from the disk surface (by the thermal pressure) 
be flung out along the open field lines. 
The matter flowing in the jet trails behind the field on a given surface. 
This generates a toroidal field component $B_\phi$
and, as a consequence, a ``magnetic torque''  
that brakes the disk, and extracts energy and angular momentum from it. 
In the acceleration process, the matter  reaches a point where 
the magnitude of the poloidal velocity equals  
the Alfv\'en poloidal velocity, or  
$|v_{\mathrm{p}}| = c_{\mathrm{A}} = |B_{\mathrm{p}}|/\sqrt{4 \pi \rho}$
($\rho$ is the mass density). 
The loci of such points constitute the so-called  Alfv\'en surface.
In the self-similar disk wind models (see, e.g., Casse \& Ferreira 
\cite{Cas00}), this surface is conical, 
and its section is also indicated in Fig. 
\ref{diskjet} with a dashed line.  
Above the Alfv\'en surface, the inertia of the matter
overcomes the magnetic forces, and the intensity of the toroidal 
field is increased substantially. This in turn generates a magnetic 
force (``hoop stress'') directed toward
the axis, that can  collimate  the flow.

The Alfv\'en surface lies at a few AU above  
the disk, a region that is still not resolved by observations.
The datapoints we obtain in the first 0\farcs2 - 0\farcs8 from the star,
however, correspond to the region immediately above, and 
can be used to investigate 
the properties of the gas that has just been accelerated and
collimated. 
The steps of the adopted procedure are illustrated in what follows.
Measured and derived quantities are summarized in Tables \ref{bigtable1}
and \ref{bigtable2}, respectively.  

\subsection{Footpoint radii}

The inner edge of the ejection region in the disk is commonly identified
with  the point where the disk is truncated by the interaction with the 
stellar magnetosphere. The
corresponding disk annulus rotates at the same rate as the central object, and
we will take this ``corotation radius'' $r_{\mathrm{c}}$ 
as a fiducial value for $r_{0,\mathrm{in}}$.
For a typical T Tauri star $r_{\mathrm{c}} \sim$  0.03 AU 
(Shu et al. \cite{Shu00}).

An estimate of the outer radius $r_{0,\mathrm{ext}}$ of the ejection  region 
for the matter seen in optical lines can be derived from the combination of 
the poloidal and toroidal velocities ($v_{\mathrm{p}}, v_{\phi}$)   
we measured in  the external layer of the flow  
located at $r_{\mathrm{obs}}$ 
from the jet axis. In a magneto-centrifugal wind the
footpoint radius of the jet component located at $r_{\rm obs}$  from the 
axis with measured $v_{\phi}$ and $v_p$ 
can be estimated with the
relationship valid at large distances from the source 
provided by Anderson et al. (\cite{And03}):
\[
r_0 \approx  0.7\,\mathrm{AU} \left( {r_{\mathrm{obs}} \over 10\,\mathrm{AU}}  \right)^{2/3}   
\left( {v_\phi \over 10\,\mathrm{km}\,\mathrm{s}^{-1} }  \right)^{2/3}
\]
\begin{equation}
~~~~~~~~~~~~~~~~~~~~~~~    
\times
\left( {v_{\mathrm{p}}  \over 100\,\mathrm{km}\,\mathrm{s}^{-1}} \right)^{-4/3}   
\left( {M_\star \over 1~~ M_\odot}  \right)^{1/3},
\label{errezero}   
\end{equation} 
in which we will take $M_\star = 1.3 \pm 0.1 $ M$_\odot$ for the mass of 
RW Aur A (Woitas et al. \cite{Woi01}). 
This simple equation is valid under the condition 
$v_{\phi} \ll v_{\mathrm{p}}$, which in the context of 
disk winds is achieved 
as soon as $r > r_{\mathrm{A}}$, 
where the `Alfv\'en radius' $r_A$ is the distance from the axis  
at which the magnetic/flow
surface rooted at $r_0$ intersects the Alfv\'en surface.
We note that at the sampled distances
above the disk (50 - 100 AU), this asymptotic regime   
may not have yet been entirely reached by the 
outer streamlines of the flow.
It can be shown, however, that in the case of `light' TTS winds, like the one
treated here, the results given by the simplified formula above 
do not differ by more than a few percent from those obtained by a 
rigorous treatment of the conservation equations  
(see Anderson et al. (\cite{And03})).
 
According to the expression above, 
in the red lobe, and for $r_{\mathrm{obs}}  = 30 $ AU (slits $S7-S1$),
one has $v_{p,\mathrm{Red}} \sim 130 ~\pm~ 10 $ 
km s$^{-1}$ (Paper~I). For the  
toroidal velocity in the first 
0\farcs2 - 0\farcs8 from the source we take 
$v_{\phi,\mathrm{Red}} = 16.8 ~\pm ~5.3 $  km s$^{-1}$, 
from the average of the significant
datapoints corresponding to the forbidden lines 
(H$\alpha$ points are excluded because of the possible 
contamination with light coming from 
the star and accretion processes, and larger projection effects). 
From these values one obtains $r_{0,\mathrm{Red}} = 1.58 ~\pm~ 0.36 $ AU.

In the blue lobe the forbidden lines are quite faint 
in the $S1$ and $S7$ slits, but the needed velocity shifts
can be extracted from the slit pair $S6-S2$. Taking the average of the 
significant values measured  within 0\farcs8  from the star  one  obtains
$v_{\phi,\mathrm{Blue}} = 14.9~ \pm~ 4.5 $ km s$^{-1}$ at 20~AU from the axis. 
With  $v_{p,\mathrm{Blue}} \sim 260 \pm 10 $ km s$^{-1}$,
the origin of this portion of the flow appears to be located at
$r_{0,\mathrm{Blue}} = 0.44 ~\pm~ 0.1 $ AU. 
The values for both the red and blue lobes 
are in  good agreement with analogous determinations
by Coffey et al. (\cite{Cof04}). We note that according to  
Pesenti et al. (\cite{Pes04}) the 
toroidal velocity derived from measurements of the full velocity profile 
(as in our case) at $r_{\mathrm{obs}} =~$20~AU from the
axis may be underestimated by a maximum of 
15\% because of projection effects. A 15\% correction to the 
toroidal velocity for the blue lobe, however, is
smaller than our accuracy and will not be considered in the following.
  
It should also be remembered that the above calculation 
strictly gives only the footpoint of the flow surface 
for which the rotational velocity could be measured, 
and {\em not} the outer radius of the whole ejection region. 
For example,  Takami et al. (\cite{Tak04}) report about 
the discovery of a cold and slow wind component 
emitting in H$_2$ lines, that surrounds the 
base of the optical jet from DG Tau. Such a component is probably 
anchored at larger footpoint radii than the optical component.
Thus, the derived footpoint values may be conservatively 
considered as lower limits to the true extent of the launching region.

\subsection{Jet/disk angular momentum balance}

In the magneto-centrifugal approach, the angular momentum 
balance in all space can
be written in conservative form as (Casse \& Ferreira \cite{Cas00} ): 
\begin{equation}
 {\bf \nabla } \cdot  \left( \rho r v_\phi  {\bf v_p}  
- {r B_{\phi} \over 4 \pi} {\bf B_p}
 - r  {\bf T_{\mathrm{v}} }  \right)~ =~   0 , 
\label{eq:divang}
\end{equation} 
where ${\bf T}$ is the `viscous'  stress  
tensor inside the disk (Shakura \& Sunyaev \cite{Sha73}), and use has been made of 
the relation  
${\bf \nabla} \cdot r {\bf T_{\mathrm{v}}} = r ({\bf \nabla} \cdot
 {\bf T}_{\mathrm{v}}). $
Viscosity can transfer the excess angular momentum  radially
from the inner to the outer region of the disk, but it is very difficult to 
quantify its importance
(see the discussion in Casse \& Ferreira \cite{Cas00}).
Our rotation measurements, however, allow us in principle to 
give rough estimates of the various contributions as follows.

We integrate Eq. \ref{eq:divang} over the volume delimited laterally 
by the magnetic surfaces passing through $r_{0,\mathrm{in}}$ and 
$ r_{0,\mathrm{ext}} $,
and above and below the disk by two surfaces, $S_\mathrm{Red}$ and
 $S_{\mathrm{Blue}}$, 
parallel to the disk plane and located in the 
observed regions at $\sim ~ \pm$  0\farcs4 AU from the source.  
We are thus led to calculate four surface integrals.
By definition ${\bf B_{\mathrm{p}}} \cdot {\bf n} = 0$ along the magnetic 
surfaces in all space,  and
in the jet ${\bf T}_{\mathrm{v}} = 0 $ (ideal conditions), and
${\bf v_{\mathrm{p}}} \parallel  {\bf B_{\mathrm{p}}}$. Thus 
the integral over the lateral surfaces reduces to the fluxes through the 
annuli at  $r_{0,\mathrm{in}}$ and   $r_{0,\mathrm{ext}}$. 
Since in the disk the poloidal velocity is equal to the accretion velocity,
the sum of the kinetic component of the flux through the annuli
under consideration is

\begin{equation}
~\dot{M}_{\mathrm{in}}~ r_{0,\mathrm{i}n}~ v_{\mathrm{K,in}} - 
\dot{M}_{\mathrm{ext}}~ r_{0,\mathrm{ext}}~ v_{\mathrm{K,ext}} = - \dot{L}_{\mathrm{disk}} ,  
\end{equation}

where $v_{\mathrm{K}} = (G M_*/r_0)^{1/2} $
is the Keplerian velocity, and $\dot{M}_{\mathrm{in}}$, 
$\dot{M}_{\mathrm{ext}}$ are the mass accretion rates in the disk 
through the magnetic surfaces located at $r_{\mathrm{0,in}}$,
$r_{\mathrm{0,ext}}$ respectively (cf. Fig.\,6).
Thus $\dot{L}_{\mathrm{disk}}$ represents the angular momentum that has to 
be extracted from the disk zone under consideration 
per unit time to allow for mass accretion. 
The component of the integral over the same surfaces due to viscosity
will be indicated by  $\dot{L}_{\mathrm{T}}$, while 
the  angular momentum fluxes in the jet lobes, $\dot{L}_{\mathrm{jet,R}}$ and
$\dot{L}_{\mathrm{jet,B}}$ result from the calculation of 
the integrals over $S_{\mathrm{Blue}}$ and  $S_{\mathrm{Red}}$. 
In summary, one reduces to:

\begin{equation} 
\dot{L}_{\mathrm{disk}}~=~\dot{L}_{\mathrm{T}} ~+ ~\dot{L}_{\mathrm{jet,Red}}~ +~ \dot{L}_{\mathrm{jet,Blue}}.
\end{equation}
 
Our rotation measures can be used to give estimates of the angular momentum 
loss rate  in the disk,   
$\dot{L}_{\mathrm{disk}} $, and of the angular momentum extracted by the jet
per unit time,  
$\dot{L}_{\mathrm{jet}}$.
A comparison then  between these quantities  can 
give us a rough idea of the role played by the outflow, and, in turn, by  
viscosity, in the angular momentum balance.

To calculate $\dot{L}_{\mathrm{disk}} $ we note that 
mass conservation leads to

\begin{equation}
\dot{M}_{\mathrm{ext}} = \dot{M}_{\mathrm{jet,Red}} + \dot{M}_{\mathrm{jet,Blue}} + \dot{M}_{\mathrm{in}}. 
\end{equation}

Referring to Fig. \ref{diskjet}, we assume that  all  the mass flowing
through the annulus at $r_{\mathrm{0,in}}$ is
deposited onto the star. Thus
a good estimate for the  inner mass flux is
$\dot{M}_{\mathrm{in}} = 1.6~10^{-6}$ M$_\odot$ yr$^{-1}$, that was derived 
by Hartigan et al. (\cite{Har95}) from measurements of the veiling of photospheric 
lines caused by the accretion shock. 
The mass flux at the jet base in the two lobes, 
$\dot{M}_{\mathrm{jet,Red}}$ and $\dot{M}_{\mathrm{jet,Blue}}$, has been  
estimated in Paper~I to be 
0.6 and  1.1~10$^{-7}$ M$_\odot$ yr$^{-1}$, 
respectively. 
We thus obtain $\dot{M}_{\mathrm{ext}} = 1.77 ~10^{-6}$ M$_\odot$ yr$^{-1}$.
Taking conservatively  
$r_{\mathrm{0,ext}}=1.58 \pm 0.36$ AU, 
and $r_{\mathrm{0,in}} \geq 0.03$ AU, one has
\begin{equation} 
\dot{L}_{\mathrm{disk}} \leq 6.6 \pm 1.5\;10^{-5}\,\mathrm{M_\odot\,yr^{-1}\, AU\,km\,s^{-1}}. 
\end{equation}

We will now attempt an estimate of 
$ \dot{L}_{\mathrm{jet, Red}}$ and $ \dot{L}_{\mathrm{jet, Blue}}$. 
Since in the jet  
$\rho {\bf v_{\mathrm{p}}} = k(a) {\bf B_{\mathrm{p}}}$, 
this term can be written, 
for each lobe, as:  
\begin{equation}
\dot{L}_{\mathrm{jet}} ~ = ~ \int_{S} \left( r v_{\mathrm{\phi}} 
- {r B_{\phi} \over 4 \pi k} \right) \rho {\bf v_{\mathrm{p}}} \cdot {\bf n}~ dS.
\label{eq:integrjet} 
\end{equation} 
The MHD wind theory 
prescribes that the quantity in parentheses, which represents 
the total (kinetic plus magnetic) specific angular momentum,
is also constant along the magnetic surfaces :
\begin{equation}
l_{\mathrm{jet}}(a) =  ( r v_\phi  
- r B_{\phi} / 4 \pi k) = r_A^2 \Omega_0 
\end{equation}
where $\Omega_0 =v_K /r_0$ is the angular speed of the disk at the 
base of the considered surface (see K\"onigl \& Pudritz \cite{Koe00}).
Equation \ref{eq:integrjet} can thus be transformed to 
\begin{equation}
\dot{L}_{\mathrm{jet}} ~ = ~ \int_{S} \left( {r_A \over r_0} 
\right)^2  r_0  v_{\mathrm{K}}  \rho {\bf v_{\mathrm{p}}} \cdot {\bf n}~ dS.
\label{eq:integrjet2} 
\end{equation} 
The quantity  $r_A \over r_0$ represents the `magnetic lever arm' of
the flow, and it varies between 2 and 5 for typical flow parameters 
(see, e.g., K\"onigl \& Pudritz \cite{Koe00}, 
Anderson et al. \cite{And03}, 
Pesenti et al. \cite{Pes04}). 
Assuming then that the flow surfaces are well-behaved and regularly nested, 
an upper limit to the angular momentum flux transported by the jet 
can be estimated from the measured jet mass flux rates,  the 
derived values of $r_0$ and  $v_{\mathrm{K}}$ 
at the outer footpoints (reported in Tables \ref{bigtable1}
and \ref{bigtable2}), and assuming a value for the magnetic lever arm.
For example with $(r_A/r_0)^2 = 10 $  one obtains 
$|\dot{L}_{\mathrm{jet, Red}}| \leq 2.6 \pm 0.8~ 10^{-5}$ 
M$_\odot$ yr$^{-1}$ AU km s$^{-1}$ and 
$|\dot{L}_{\mathrm{jet, Blue}}| \leq 2.5 \pm 1.2~ 10^{-5}$ 
M$_\odot$ yr$^{-1}$ AU km s$^{-1}$. When
summed, this amounts already to about 80\% of the excess angular 
momentum, $\dot{L}_{\mathrm{disk}}$,
to be extracted from the portion of the disk from which the 
observed flow originates. 

An interesting  lower limit to the angular momentum transported by  the jet 
can instead be given  for the
self-similar disk wind model of Ferreira \& Pelletier (\cite{FerPer93}).
Before deriving this 
we note that  the 3-D geometry
of the {\bf B} and {\bf v}  
vectors in the magneto-centrifugal scenario is such that 
if the system rotates anticlockwise 
looking down the $+z$ axis ($v_{\mathrm{\phi}}, v_{\mathrm{K}}  >  0$),
then $B_{\phi} B_{\mathrm{p}} / \rho  v_{\mathrm{p}}  = B_{\phi} / k(a)
\leq 0 $ in all 
space (i.e., in both the $z>0$ and $z<0$ domains), 
while the opposite holds for clockwise rotation. 
Therefore,  for every possible 
configuration of the system the kinetic and magnetic terms in the specific 
angular momentum have opposite signs, or they add up in absolute value. 
Thus, for each lobe, a lower limit to $|\dot{L}_{\mathrm{jet}}|$
is set assuming that the magnetic contribution is negligible.
The integral in Equation \ref{eq:integrjet} can be computed 
exactly for the model of Ferreira \& Pelletier (\cite{FerPer93}). 
For this model  $v_{\mathrm{\phi}}, v_z \propto r^{-1/2} $ and  
$\rho  \propto r^{(\xi-3/2)}$,  
where $\xi$ is the ejection index of the jet (see next section). 
In this case Eq. \ref{eq:integrjet} reads:
\[
|\dot{L}_{jet} | \geq ~ 
~ \int_{S}  r v_\phi \rho {\bf v_p} \cdot {\bf n}~ dS
~\approx~
v_{\phi}(r_{\mathrm{obs}})~r_{\mathrm{obs}}^{1/2}~
\]
\begin{equation}
~~~~~~~~~~~~~~~
\times ~{\xi \over \xi+1/2}~
{r_{\mathrm{obs}}^{(\xi+1/2)} - r_{\mathrm{in}}^{(\xi+1/2)} 
\over r_{\mathrm{obs}}^{\xi} - r_{\mathrm{in}}^{\xi}}
~\dot{M}_{\mathrm{jet}},
\label{eq:Ljet} 
\end{equation}
where the integral is calculated between the 
radius of the axial inner hole of the jet $r_{\mathrm{in}}$, that is about 
5 AU at 0\farcs4 from the star (Cabrit et al. \cite{Cab99})  and the radius 
of the observed external layer, $r_{\mathrm{obs}}$, which is 30~AU and 
20~AU for the red and blue lobes, respectively.
With  $\xi = 0.040$, a value similar to the one that will be 
derived for this jet in the next section, we obtain 
$|\dot{L}_{\mathrm{jet, Red}}| \geq 2.4 \pm 0.8~ 10^{-5}$ 
M$_\odot$ yr$^{-1}$ AU km s$^{-1}$ and 
$|\dot{L}_{\mathrm{jet, Blue}}| \geq 2.0 \pm 0.9~ 10^{-5}$ 
M$_\odot$ yr$^{-1}$ AU km s$^{-1}$.
Again summing these quantities up the total amounts to  
about two thirds of the angular momentum $\dot{L}_{\mathrm{disk}}$, 
that has to be extracted from the 
examined portion of the disk per unit time.

The rough estimates above indicate that the mechanism generating 
the jet may indeed be capable 
of braking the disk with high efficiency.  
In the rest of this discussion we will assume that 
in the system under study  all of the excess angular 
momentum is carried away by the wind, or, equivalently, that 
in the disk the magnetic torque is much larger than the viscous torque.

\subsection{Magnetic lever arm  and ejection index}

Following  the disk-wind formulation by Ferreira (\cite{Fer97}, 
\cite{Fer02}), in a jet/disk structure like the one considered, 
the mass accretion flux through the disk, $\dot{M}_{\mathrm{acc}} $, 
varies with distance from the star as  
$\dot{M}_{\mathrm{acc}} = Const \times r_0^{\xi}$. The exponent $\xi$, that 
measures the local ejection efficiency, is called 
`ejection index', and 
regulates many of the physical properties of the wind. 
Under the assumption of dominant magnetic torque (see above),
$\xi$ is simply related to 
$r_{\mathrm{A}}/r_0$ by the expression: 
$(r_{\mathrm{A}}/r_0)^2 = \lambda \sim 1 + 1/2\xi$.

According to the disk-wind  models 
0.005 $< \xi < $ 0.5, 
where the upper limit applies to self-similar  
solutions with  sub-Alfvenic heating (warm solutions, 
see Casse \& Ferreira \cite{Cas00}).
It is interesting to find limiting values for this parameter 
in our case. 

A lower limit for $\xi$ is set by the conservation 
of mass in the disk/jet  system.
The external mass flux can  be expressed through $\xi$ and 
$\dot{M}_{\mathrm{in}}$ as

\begin{equation}
\dot{M}_{\mathrm{ext}} = 
\dot{M}_{\mathrm{in}} (r_{\mathrm{0,ext}}/ r_{\mathrm{0,in}})^{\xi}. 
\end{equation}

Since $r_{\mathrm{0,in}}$ cannot be lower than the 
corotation radius $r_{\mathrm{c}}$, a lower limit for $\xi$ is given by: 
\begin{equation}
\xi \geq {\ln ( 1 + ( \dot{M}_{\mathrm{jet, Red}} + 
\dot{M}_{\mathrm{jet, Blue}})/\dot{M}_{\mathrm{in}}) \over 
\ln (r_{\mathrm{0,ext}} / r_{\mathrm{c}})  }
\end{equation} 
Taking for $r_{\mathrm{0,ext}}$ the value of $r_0$ derived for the red lobe 
(1.58 AU) 
one obtains $\xi_{\mathrm{min}}$ = 0.025 $\pm$ 0.005.
If instead we take $r_{\mathrm{0,ext}} = 0.44$ AU, that is the
lateral footpoint  value for the blue lobe, one obtains  
$\xi_{\mathrm{min}} =  0.037 \pm 0.005 $.  
The value of $\xi_{min}$  may further differ 
if one considers that the full extent  
of the accretion/ejection region may be larger, due to the 
possible presence of a surrounding cold flow  
not visible at optical wavelengths (such as a H$_2$ wind, 
cf. Takami et al. \cite{Tak04}). This  would imply not only a larger 
$r_{\mathrm{0,ext}}$, but also a larger $\dot{M}_{\mathrm{jet}}$, 
as one would have to consider also the contribution   
of such flow to the mass outflow rate.  
Both these quantities, however, 
are still undetermined for the case under study.

An upper limit for $\xi $ follows from the conservation of angular 
momentum in the jet region. For dominant magnetic torque
the constant of motion can be expressed through $\lambda$  or $\xi$  
in terms of the  Keplerian rotation velocity  at the footpoint: 
%
\[
|l_{\mathrm{jet}} (a) |= \left| r \left(v_\phi  -  {B_{\phi}  \over 4 \pi k } \right) \right| =
\lambda~ v_{\mathrm{K}} r_0 
\]
\begin{equation}
~~~~~~~~~~~~~=
\left(1 + {1 \over 2 \xi } \right )~
(G M_*)^{1/2}~ r_0^{1/2}.   
\label{eq:angmom}
\end{equation}
Now, as described in the previous 
paragraph, the kinetic and magnetic terms in $l$ 
always have opposite signs, 
so the condition $|B_{\phi}| \geq 0$ imposes 
an upper limit for $\xi$.   
For the red lobe, using $r_0 =$ 1.58 AU, one finds 
$\xi_{\mathrm{max,Red}} = 0.046 \pm 0.011$,  while for the 
blue lobe, with  $r_0 =$ 0.44 AU, one obtains 
$\xi_{\mathrm{max,Blue}} = 0.041 \pm 0.010$.

In summary, for the red lobe of the 
RW Aur jet 0.025 $\pm 0.005 \leq \xi \leq  0.046
\pm 0.011 $, or  $12 \pm 3  \leq \lambda \leq 21 \pm 4$.
At the same time, in the blue lobe one finds 
 0.037 $\pm 0.005 \leq \xi \leq  0.041
\pm 0.011 $, or  $13 \pm 3  \leq \lambda \leq 15 \pm 4$.
The derived values for $\xi$ are thus 
in the lower part of the range admitted by the model.
Nevertheless, ``cold'' wind ejection (enthalpy negligible 
with respect to the 
kinetic energy at the footpoint) requires $\xi \la 0.01$ 
(Casse \& Ferreira \cite{Cas00}). 
Thus our results are consistent with the idea
that some sort of heating is provided at the base of the jet
(i.\,e. the wind is ``warm''), for example
by a hot disk corona. 
The fact that an extra-heating seems to be required 
at this location has also been
noticed by Garcia et al. (\cite{Gar01a}) from the analysis 
of terminal poloidal velocities and jet total densities, and by Pesenti et al. 
(\cite{Pes04}) from the analysis of rotation signatures in the DG Tau jet.
If the heating is produced by an active disk, 
however, a certain  amount of viscous dissipation must be allowed. 
Alternatively, the heating at the jet footpoints might be due to
ambipolar diffusion and/or X-ray irradiation. See 
Garcia et al. (\cite{Gar01b}) and Shang et al. (\cite{Sha02}) 
for a discussion of these processes respectively.   

\subsection{Properties of the magnetic field}

A quantity that can be derived directly from our measurements 
is the ratio  $B_{\phi} / B_p$ between the 
toroidal and poloidal components of the magnetic 
field vector at the location of the observations. 
This quantity indicates how much the lines of force are wrapped 
on a given magnetic surface. In the magneto-centrifugal scenario, the 
collimation of the jet is thought to arise from the hoop stress 
generated by the increase of the toroidal component of the magnetic field 
after the gas has passed through the Alfv\`en surface (see, e.g. K\"onigl
\& Pudritz \cite{Koe00}). 
It is thus interesting to obtain from the observations an indication 
of the magnetic field configuration in the region just above the 
collimation zone.

The ratio  $B_{\phi} / B_p$
can be  derived using a further  conservation 
law of general disk-wind theory (see K\"onigl \& Pudritz \cite{Koe00}, 
Anderson et al. \cite{And03}):  
\begin{equation}
\left(v_{\phi} - {B_{\phi} v_p \over B_p} \right) / r  = \Omega_0,  
\label{eq:pitch}
\end{equation} 
where $\Omega_0$ is the disk angular velocity at the footpoint. 
Introducing the values observed for $v_{\phi}$, $v_p$ at $r= r_{\mathrm{obs}}$
in the two lobes, and using the corresponding $r_0$ to calculate $\Omega_0$,
we find  $B_{\phi}/ B_p$ = 
3.8 $\pm$ 1.1 in the red lobe, at 30 AU from the axis and about 80 AU
from the disk. In the blue lobe, one finds  $B_{\phi}/ B_p$ =
  -8.9 $\pm$ 2.7 at 20 AU from the axis and at about 100 AU
above the disk (the signs refer to a cylindrical coordinate system with 
the positive $z$ axis oriented along the blue lobe). 

If the poloidal magnetic field is almost entirely projected along 
the $z$ direction, these ratios would 
correspond to an inclination angle of the 
magnetic line of force with respect to the disk plane 
of about 15 degrees for the red lobe, and only 6 -- 7 degrees for 
the blue lobe. The dominance of the toroidal component 
of the field appears  to confirm  that the flow is  
magnetically collimated, as prescribed by the 
magneto-centrifugal models.
Also, the well-known asymmetry between the lobes of this jet is reflected 
in the  magnetic configuration, as the field  
turns out to be more tightly wrapped in the blue lobe.

%

\begin{table*}[!ht] \footnotesize
\caption{{\bf Measured physical quantities at the jet base}}
\label{bigtable1}
\begin{tabular}{lcccccc} \hline
\hline
 & & & & & & \\
Jet lobe    & $r_{\mathrm{obs}}$  &  $v_{\phi}^{\rm a,b}$ & $v_p^{\rm b}$  & $M_{\star}^{\rm c}$  & $\dot{M}_{\mathrm{in}}^{\rm d}$ & $\dot{M}_{\mathrm{jet}}^{\rm e}$\\ 
& (AU) & (km s$^{-1}$) & (km s$^{-1}$)  & (M$_{\odot}$) & (10$^{-6}$ M$_{\odot}$ yr$^{-1}$) & (10$^{-6}$ M$_{\odot}$ yr$^{-1}$) \\
 & & & & & & \\
\hline
&&&&&&\\
{\bf Redshifted}  & 30 $\pm$ 3 & 16.8 $\pm$ 5.3  & -130 $\pm$ 10 &  &  & 0.06 $\pm$ 0.01 \\
&&&& 1.3 $\pm$ 0.1 & 1.6 $\pm$ 0.1 &\\
{\bf Blueshifted} & 20 $\pm$ 3 & 14.9 $\pm$ 4.5  & 260 $\pm$ 10 & & & 0.11 $\pm$ 0.01\\
&&&&&&\\
\hline 
&&&&&&\\
\end{tabular}
\begin{list}{}{}
\item[$^{\mathrm{a}}$] Average of the significant values obtained from the FELs 
between 0\farcs2 and 0\farcs6 (0\farcs8) for the red (blue) lobe. 
\item[$^{\mathrm{b}}$]
The vector components are in 
a cylindrical coordinate system centred on the star and 
with positive $z$ axis oriented 
along the blue lobe (South-East in Fig. 1).
\item[$^{\mathrm{c}}$] From Woitas et al., \cite{Woi01}.
\item[$^{\mathrm{d}}$] From Hartigan et al. \cite{Har95}.
\item[$^{\mathrm{e}}$] From Paper~I.

\end{list}
\end{table*}
\vspace{1cm}

\begin{table*}[!ht] \footnotesize
\caption{{\bf Derived physical quantities}}
\label{bigtable2}
\begin{tabular}{lccccccc} \hline
\hline
&&&&&&&\\
Jet lobe & $r_0^{\rm a}$ & $v_K^{\rm b}$ 
& $|\dot{L}_{\mathrm{jet,min}}|^{\rm c}$ &  $\dot{L}_{\mathrm{disk,max}}^{\rm d}$ &  $\xi^{\rm e}$ & $\lambda^{\rm e}$ & $B_{\phi}/B_p^{\rm f}$   \\ 
  & (AU) & (km s$^{-1}$) & (M$_{\odot}$yr$^{-1}$AU\,km/s)  & (M$_{\odot}$yr$^{-1}$AU\,km/s) &&&  \\
\hline 
&&&&&&&\\
{\bf Red} & 1.58 $\pm$ 0.36 &  27.0 $\pm$ 3.2 & 2.4 $\pm$ 0.8  10$^{-5}$ &  &

  0.025 $\leq \xi \leq $ 0.046 & 12 $\leq \lambda \leq $ 21 &  3.8 $\pm$ 1.1  \\
& & & & 6.6 $\pm$ 1.5 10$^{-5}$& & & \\
{\bf Blue} & 0.44  $\pm$ 0.10  & 51.2 $\pm$ 6.2 & 
2.0 $\pm $ 0.9 10$^{-5}$ & &  
0.037 $\leq \xi \leq $  0.041 & 13 $\leq \lambda \leq $ 15 & -8.9 $\pm$ 2.7   \\
&&&&&&&\\
\hline
\end{tabular}
\begin{list}{}{}
\item[$^{\mathrm{a}}$] Derived with the method in Anderson et al. \cite{And03}.\item[$^{\mathrm{b}}$] Keplerian velocity at the footpoints $r_0$. 
\item[$^{\mathrm{c}}$] Imposing that $|B_{\phi}|=0$ at the location of the observations and adopting the scalings in Ferreira \& Pelletier 1993 (see text).
\item[$^{\mathrm{d}}$] Considering only the portion of the disk from where the outflow seen at optical wavelengths originates. 
\item[$^{\mathrm{e}}$] Assuming dominant magnetic torque; accuracy 20 - 22 \%.
\item[$^{\mathrm{f}}$] At about 80 AU above the disk and 30 (20) AU from the 
axis for the red (blue) lobe. Coordinate system as in Table \ref{bigtable1}.

\end{list}
\end{table*}

\section{Conclusions} \label{summ}

In this paper we  have described a new combined observational/theoretical
study of the rotation properties of the bipolar jet from RW Aur.

We have found rotation signatures in a set of spectra taken 
with HST/STIS  with multiple slits
oriented parallel to the flow axis, using 
techniques similar to the ones  adopted in our first rotation
study (of the DG Tauri jet, Bacciotti et al. \cite{Bac02}).
We analyse the first 300 AU of the jet (1\farcs5 at the distance of RW Aur), 
applying an updated version of the correction routines for uneven
slit illumination. We find that both lobes rotate in 
the same direction (i.e. with different helicities), with toroidal velocities
in the range 5 - 30 km s$^{-1}$ 
at 20 and 30 AU from the symmetry axis in the blue and red 
lobes, respectively. 
The sense of rotation is anti-clockwise looking from the tip of the blue lobe 
(P.A. 130$^{\circ}$  North to East) down to the star. 
These results are confirmed by other HST/STIS observations from our group 
(with the slit placed transverse to the jet axis)
presented in Coffey et al.\, (\cite{Cof04}).
Rotation is more pronounced in the data from the outermost slit pair,  
as expected because of  projection effects along the line of sight
(Pesenti et al. \cite{Pes04}). Also, rotation signatures are more evident 
in [OI] and [NII] lines than  in  H$\alpha$ and [SII] lines. 
We interpret this result as due to the fact that [OI] and [NII] emission
traces regions closer to the jet axis  than H$\alpha$ and [SII], 
and hence at a given position of the slit the jet is observed closer to its 
border in [OI] and [NII], which reduces the confusion due to projection. 

The observed rotation favours widely known magneto-centrifugal
models for the jet generation. 
Following  the formulation 
in Anderson et al. (\cite{And03}), 
the  derived  toroidal and poloidal velocities imply that the
flow surfaces of the  redshifted lobe
observed at 30 AU from the axis  
are rooted in the disk at about 1.6 AU from the star.
The blueshifted flow observed at 20 AU from the axis arises 
instead from a region in the disk at about 0.4 -- 0.5 AU from the star. 

Making use of general principles underlying the disk-wind models,
we have also derived other parameters useful to constrain 
the properties of the RW Aur accretion/ejection structure.
We have estimated upper limits for $\dot{L}_{\mathrm{jet}}$, 
the angular momentum transported by the visible jet lobes, 
and a lower limit for the same quantity in the special 
case of the self-similar disk-wind model of
Ferreira \& Pelletier (\cite{FerPer93}). 
We compare these values with the angular 
momentum, $\dot{L}_{\mathrm{disk}}$, that the region of the 
disk from which the visible  outflow originates 
has to lose per unit time in order to  
accrete at the observed rate. We conclude that 
the jet is capable of extracting a consistent 
fraction (two thirds or more)
of the excess angular momentum present in the disk.

Assuming moreover that {\em all} the excess angular momentum is
carried away by the jet, we have also estimated  the magnetic lever arm   
(expressed by the ratio $r_A/r_0$ between  the Alfv\'en
and footpoint radii) for the  
self-similar disk wind model of Ferreira \& Pelletier (\cite{FerPer93}).
We found this quantity to be in the range 3.6 - 3.9 for the blueshited lobe, 
and in the range 3.5 - 4.6 in the redshifted lobe (accuracy 20 - 25 \%). 
Alternatively,  the value of the ejection index 
$\xi = d\ln ( \dot{M}_{\mathrm{acc}} ) / d r $ varies from  
0.025 to  0.046  in the red lobe, and from 0.037 to 0.041 for the
blue lobe (with the same accuracy). 
We caution that our  determination of the magnetic
lever arm (or ejection index)  may be affected by 
the poor knowledge of the global extension of the launching region. In fact 
a component of the flow too cold to be visible at optical wavelengths
may surround the observed jet.
Nevertheless, the values determined for the optical component
are in the range predicted by MHD models, and they also suggest that some 
heating is provided externally at the base of the flow.
The nature of such  heating, however, remains to be identified. 

Finally, we have used our rotation measurement to derive 
information about the  spatial configuration  of the magnetic field
in the examined region.
In particular, using well-known conservation laws 
of the MHD disk wind theory we have derived the ratio $B_{\phi}/B_p $
between the toroidal and poloidal components of the magnetic field
at the observed locations in  both lobes of the bipolar jet. 
We obtained  $B_{\phi}/B_p  = 3.8 \pm 1.1$ for the red lobe at 30 AU 
from the axis and about 80 AU from the disk  and 
$B_{\phi}/B_p  = -8.9 \pm 2.7$ for the blue lobe, at 20 AU 
from the axis and 100 AU from the disk 
(in cylindrical coordinates, with positive $z$ along the blue lobe).
The toroidal component of the magnetic field appears thus to be 
dominant, as expected for a magnetically collimated flow.  
In addition, the field seems to be more tightly wrapped in the blue lobe,
reflecting the well-known (but unexplained) asymmetries between the 
two lobes of this jet.

In summary, our observations and subsequent analysis
appear to confirm once more the magneto-centrifugal scenario 
for the launching of YSO jets. 
To prove conclusively that jets solve the ``angular momentum'' problem
in star formation, however, will require further detailed studies of
a larger number of jets, and possibly at higher spectral and spatial 
resolution.

\begin{acknowledgement}
J.E. and J.W. acknowledge support by the Deutsches Zentrum f\"ur
Luft- und Raumfahrt (grant number 50 OR 0009), and T.R. and D.C. 
funding from Enterprise Ireland. We would like to thank the referee,
Catherine Dougados, for a helpful and constructive report that led to a
significant improvement of this paper. We are grateful to
Jonathan Ferreira, Ralph Pudritz and Antonella Natta for useful comments.
T.R., J.E., D.C., and J.W.
wish to thank the Arcetri Observatory for hospitality
during various visits.
\end{acknowledgement}

\end{document}